# Evidence for pressure induced polarization rotation, octahedral tilting and reentrant ferroelectric phase in tetragonal $(Pb_{0.5}Bi_{0.5})(Ti_{0.5}Fe_{0.5})O_3$

Pragya Singh[1], Chandan Upadhyay[1], Zuzana Konôpková[2], Hanns-Peter Liermann[2], and Dhananjai Pandey[1]

[1]School of Materials Science and Technology, Indian Institute of Technology (Banaras Hindu University), Varanasi 221005, India

[2]PETRA III, Deutsches Elektronen-Synchrotron (DESY), 22607 Hamburg, Germany

**Abstract:** Despite the technological significance of the tetragonal $PbTiO_3$ for the piezoelectric transducer industry, its high pressure behaviour is quite controversial as two entirely different scenarios, involving pressure induced (1) morphotropic phase boundary (MPB) like structural transition with concomitant rotation of the ferroelectric polarization vector and (2) antiferrodistortive (AFD) phase transition followed by emergence of a reentrant ferroelectric phase, have been proposed in recent theoretical and experimental studies. We have attempted to address these controversies through a high resolution synchrotron x-ray diffraction study of pressure induced phase transitions in the tetragonal phase of a modified $PbTiO_3$ composition containing 50% $BiFeO_3$, where $BiFeO_3$ was added to enhance the AFD instability of $PbTiO_3$. We present here the first experimental evidence for the presence of the characteristic superlattice reflections due to an AFD transition at a moderate pressure $p_{c1}$ ~2.15 GPa in broad agreement with scenario (2), but the high pressure ferroelectric phase belongs to the monoclinic space group Cc, and not the tetragonal space group I4cm predicted under scenario (2), which permits the rotation of the ferroelectric polarization vector as per scenario (1). We show that the monoclinic distortion angle and ferroelectric polarization of the Cc phase initially decrease with increasing pressure for $p < 7$ GPa, but start increasing above $p_{c2}$ ~ 7 GPa due to an isostructural Cc-I to Cc-II transition reminiscent of $M_A(a_{pc} > b_{pc} \simeq c_{pc})$ to $M_B(a_{pc} < b_{pc} \simeq c_{pc})$ transition predicted for MPB systems. We also show that octahedral tilting provides an efficient mechanism for accommodating pressure induced volume reduction for the stabilisation of the reentrant ferroelectric phase Cc-II. Our results not only address the controversies regarding high pressure behaviour of PbTiO$_3$, but also provide an insight for designing new environmentally friendly lead-free piezoelectric compositions.

---------------------------

## I. INTRODUCTION

Pressure has been extensively used as a clean thermodynamic variable for studying a rich variety of phenomena like superconductivity [1-3], insulator-to-metal transition [4-5], ferroelectric transition [6-7], octahedral tilt transitions [8-10], magnetic transitions [11-13], quantum phase transitions [14-16], and optical band gap tuning [17-20] etc. Amongst the various functional materials, pressure induced phase transitions in ABO$_3$ type ferroelectric perovskites have received enormous attention as some of these compounds are the end members of commercial piezoelectric and capacitor compositions, with a market share of several tens of billions of dollars [21-22]. The structural phase transitions in perovskites are broadly classified as ferrodistortive (FD) and antiferrodistortive (AFD), which are driven by the softening of zone centre (q=0) and zone boundary (q≠0) optical phonon modes, respectively [23]. Ferroelectric (FE) and octahedral tilt transitions fall under the broad category of FD and AFD transitions, respectively. In a classic work on pressure induced phase transitions in perovskites, Samara and his co-workers [24] argued that the FE phase transitions are suppressed at high pressures due to rapid increase in the short range repulsive interactions as compared to the long range attractive interactions. In marked contrast, the octahedral tilt transitions are favoured at high pressures because the rotation/tilting of oxygen octahedra can easily accommodate the volume contraction through the bending of B-O-B bond angles without causing significant B-O bond length compression [25].

Recent theoretical and experimental investigations on high pressure behaviour of perovskites raise doubts about the validity the predictions by Samara and his co-workers [24]. PbTiO$_3$ (PT) is one such model compound whose high pressure behaviour has been revisited extensively in recent years, because of its technological importance for the piezoelectric industry. Wu and Cohen [26] using first principle calculations predicted FE tetragonal P4mm $\xrightarrow{\sim 9.5\ \text{GPa}}$ FE monoclinic Cm $\xrightarrow{\sim 11\ \text{GPa}}$ FE rhombohedral R3m $\xrightarrow{\sim 22\ \text{GPa}}$ PE cubic Pm$\bar{3}$m phase transitions at successively higher pressures, involving rotation of the FE polarization vector (**Ps**) from [001] in P4mm to [110] in Cm to [111] in R3m using pseudo-cubic indices. This sequence of phase transitions is reminiscent of composition induced phase transitions across the morphotropic phase boundary (MPB) in the technologically important complex piezoelectric perovskites like PbZr$_x$Ti$_{(1-x)}$O$_3$ (PZT), (1-x)PbMg$_{1/3}$Nb$_{2/3}$O$_3$-xPbTiO$_3$ (PMN-xPT) and (1-x)PbZn$_{1/3}$Nb$_{2/3}$O$_3$-xPbTiO$_3$ (PZN-xPT) [21, 27-28], where the rotation of polarization vector is believed to be responsible for maximization of the piezoelectric properties at the MPB [29-30]. Because of this analogy between pressure and composition induced phase transitions, it is believed that the study of pressure induced phase transitions has the potential for providing clues to designing new environmentally friendly lead-free piezoelectric perovskites [31-32]. In an independent first principles study, Kornev et al. [33], however, predicted a completely different scenario which does not involve rotation of the polarization vector, but is based on the AFD octahedral tilt transitions as per the following sequence: FE tetragonal P4mm $\xrightarrow{\sim 3\ \text{GPa}}$ FE tetragonal I4cm $\xrightarrow{\sim 12\ \text{GPa}}$ non-FE tetragonal I4/mcm $\xrightarrow{\sim 30\ \text{GPa}}$ FE tetragonal I4cm at successively higher pressures, where both the I4cm and I4/mcm phases contain tilted oxygen octahedra due to AFD instability induced by pressure. The appearance of the reentrant FE phase in the I4cm space group above ~30 GPa in these calculations may sound intriguing, as the FE distortion has been generally believed to disappear at high pressures [24]. In a subsequent work, Kornev and Bellaiche explained the appearance of the reentrant FE phase in terms of softening of the short range repulsive interactions through the hybridization of Ti $3d^0$ and O $2s^2$ orbitals [34]. First principles calculations by Frantti et al. [35] suggest a yet another scenario, different from Wu and Cohen [26] and Kornev et al. [33], for the high pressure phase transitions in PbTiO$_3$: FE tetragonal P4mm $\xrightarrow{\sim 9\ \text{GPa}}$ FE rhombohedral R3c $\xrightarrow{\sim 30\ \text{GPa}}$ PE rhombohedral R$\bar{3}$c, where neither the monoclinic Cm phase predicted in Ref. [26] nor the tetragonal I4cm or I4/mcm phases appear as predicted in Ref. [33-34]. More recently, Ganesh and Cohen [36] have argued that AFD distortions in PbTiO$_3$ cannot be stabilized at low pressures. They obtained a modified sequence for pressure induced phase transition in PT with the high pressure cubic phase replaced by rhombohedral phases (R3c or R$\bar{3}$c) above 25 GPa, but reaffirmed the possibility of MPB type polarization rotation at comparatively much lower pressures predicted earlier by Wu and Cohen [26].

While there is still no unanimity on the sequence of high pressure phase transitions in PbTiO$_3$ predicted theoretically, the experimental situation is no better. Ahart et al. [37] presented evidence for pressure induced phase transition from tetragonal P4mm to monoclinic Pm (at ~9.5 GPa) to another monoclinic Cm (at ~11 GPa) to rhombohedral R$\bar{3}$c/R3c (at ~22 GPa) phases using synchrotron x-ray diffraction and Raman scattering measurements at low temperatures in support of the theoretical predictions of Wu and Cohen [26]. On the other hand, Janolin et al. [38] obtained evidence for pressure induced decrease in tetragonality of PT going to an almost cubic-like phase, whose space group could not be assigned unambiguously even though Kornev and Bellaiche [34] have predicted the possibility of rhombohedral (R3m, R$\bar{3}$c and R3c) phases involving AFD octahedral tilting with and without ferroelectric distortion at intermediate pressures. Above 20 GPa, they showed that this cubic like phase transforms to another non-ferroelectric tetragonal phase in I4/mcm space group and later to ferroelectric I4cm phase above 45 GPa both involving AFD rotation of the oxygen octahedral [38]. The tetragonality was reported to increase continuously at high pressures (p >20 GPa). However, except for Raman scattering measurements, no direct evidence for the presence of superlattice reflections in SXRD studies expected for AFD transition has been presented at the moderately low pressures (~3 GPa) at which such transitions have been predicted theoretically [33-34]. Only at extremely high pressures exceeding 43 GPa [38], superlattice reflections have been reported in single crystal diffraction pattern. Cohen et al. [39], using second harmonic generation (SHG) measurements along with density functional theory calculations, have, however, questioned the existence of the ultrahigh pressure reentrant ferroelectric phase proposed by Janolin et al. [38] and Kornev et al. [33].

It is evident from the foregoing that there is a lot of controversy regarding phase transition behaviour of $PbTiO_3$ at high pressures. These controversies can be summarized as follows: (i) Is there a pressure induced AFD type phase transition at moderate pressures with clear evidence for the presence of superlattice peaks in the SXRD patterns? (ii) Can pressure induce morphotropic phase boundary like transition from tetragonal to rhombohedral phase via intermediate monoclinic phases with concomitant rotation of the ferroelectric polarization vector? (iii) What is the exact nature (FE or non-FE) of the intermediate cubic-like phase? (iv) Is there a reentrant ferroelectric phase at high pressures whose polarization increases with pressure?

The present investigation is undertaken to address these controversies using a tetragonal composition of the solid solution system $(1-x)PbTiO_3-xBiFeO_3$ with x = 0.50 (PTBF50). In this system, it is expected that the soft R (q = ½ ½ ½) branch of phonon spectrum of $PbTiO_3$ [40] would become more unstable due to $BiFeO_3$ substitution which should in turn enhance the intensity of the AFD superlattice peaks in the SXRD patterns [41-42]. Using high resolution synchrotron x-ray diffraction measurements, we present here the first experimental evidence for characteristic superlattice reflections due to AFD transition at a critical pressure $p_c$ ~2.15 GPa, in broad agreement with predictions of Kornev et al. for $PbTiO_3$ [33-34]. However, this high pressure phase does not belong to ferroelectric I4cm space group, but to ferroelectric monoclinic Cc space group, which suggests that the ferroelectric polarization vector indeed rotates above this critical pressure as expected for a morphotropic phase boundary like transition proposed by Wu and Cohen [26]. We also show that the monoclinic distortion angle and FE polarization of the Cc phase decrease with increasing pressure up to ~7 GPa as expected on the basis of the predictions by Samara and his co-workers [24]. However, we find that this low pressure Cc phase undergoes an isostructural phase transition to another Cc phase around 7 GPa in which both the monoclinic distortions and ferroelectric polarization increase with pressure, pointing towards the existence of a reentrant ferroelectric phase as predicted by Kornev et al. [33-34]. We show that this anomalous increase in polarization is facilitated by increase in the oxygen octahedral tilting that accommodates compression at high pressures in PTBF50. We also show that the intermediate cubic-like phase remains ferroelectric in the same space group and only represents a crossover regime from one monoclinic phase to another monoclinic phase through an $M_A(a_{pc}>b_{pc}\simeq c_{pc})$ to $M_B(a_{pc}<b_{pc}\simeq c_{pc})$ (see Ref. [43]) like isostructural phase transition.

## II. EXPERIMENTAL DETAILS

Polycrystalline samples of $0.50PbTiO_3$-$0.50BiFeO_3$ (PTBF50) were synthesized by sol-gel method, details of which can be found elsewhere [44]. Pressure dependent synchrotron x-ray diffraction (SXRD) data was collected at P02.2 beamline of PETRA III (Germany) using a wavelength of 0.29135 Å (42.8 keV photon energy) [45]. The diffraction measurements were performed using a symmetric piston cylinder type diamond anvil cell of culet size 400 μm and a pre-indented rhenium gasket of 36 μm width with sample hole size of 200 μm. Helium was used as the pressure transmitting medium. The pressure calibration was done by fluorescence method based on the shifts of the two characteristic ruby laser lines. The sample to detector distance was kept at ~900 mm and a CsI bonded amorphous silicon flat panel 2D detector XRD1621 from Perkin Elmer was used to collect the SXRD data. The 2D data was converted to 1D (intensity versus 2θ) form using FIT2D software, while the profile matching analysis and structure refinement was carried out by LeBail and Rietveld methods using FULLPROF package [46].

### III. RESULTS AND DISCUSSIONS

The room temperature crystal structure of PTBF50 under ambient pressure was found to be tetragonal in P4mm space group in agreement with the literature [47-48]. Fig. 1 shows the evolution of SXRD profiles of $(200)_{pc}$, $(220)_{pc}$ and $(222)_{pc}$, pseudo-cubic (pc) perovskite reflections of PTBF50 with increasing pressure at ambient temperature. The tetragonal (T) structure of the PTBF50 under ambient conditions is confirmed by the doublet nature of $(200)_{pc}$ {i.e., $(002)_T$ and $(200)_T$ (where the subscript T stands for tetragonal unit cell)} and $(220)_{pc}$ {i.e., $(202)_T$ and $(220)_T$} reflections and singlet nature of $(222)_{pc}$ reflection. With increasing pressure, the $(002)_T$ and $(200)_T$ as well as $(202)_T$ and $(220)_T$ peaks of the tetragonal phase approach towards each other which implies that the tetragonality ($\eta = c/a -1$) is decreasing with increasing pressure as has been observed by most workers in pure $PbTiO_3$ also [38,49]. This is the expected behaviour as per Samara's criterion which predicts gradual suppression of ferroelectric distortion (i.e., tetragonality in the present case) with increasing pressure [24]. On increasing the pressure to ~2.15 GPa, the $(222)_{pc}$ reflection, which was a singlet in the tetragonal phase, shows splitting. Further the pair of reflections $(002)_T$ and $(200)_T$ of the tetragonal phase overlap with each other giving rise to a singlet like appearance.

In addition, the $(220)_{pc}$ peak, which although remains a doublet even at ~2.15 GPa, the separation between the two peaks is now drastically reduced and the weaker reflection of this doublet occurs on the lower 2θ side of the stronger intensity peak. This is in marked contrast to the tetragonal phase for which the weaker $(220)_T$ reflection occurs on the higher 2θ side of the stronger $(202)_T$ peak. All these features indicate that a pressure induced structural phase transition has taken place in PTBF50 between 1.26 GPa and 2.15 GPa. More interestingly, weak superlattice reflections, which are indexed as $(3/2\ 1/2\ 1/2)_{pc}$ and $(3/2\ 3/2\ 1/2)_{pc}$ also appear at $p_c$ ~2.15 GPa as shown in Fig. 2(b). The observation of these superlattice reflections with fractional indices suggests that the unit cell is doubled. With respect to a doubled pseudo-cubic cell, the superlattice peaks are indexed as $(311)_{pc}$, and $(331)_{pc}$. The odd-odd-odd (ooo) nature of the superlattice peaks implies anti-phase tilting of the neighbouring oxygen octahedra as per Glazer's criterion [50-51]. Such octahedral tilt transitions are known to be driven by phonon instability at the R-point (q = ½ ½ ½) of the cubic Brillouin zone and this transition is, therefore, of AFD type [23]. First principles calculations on $PbTiO_3$ have predicted AFD transition at $p_c$ ~3 GPa [33]. However, the corresponding superlattice reflections have not been observed until the pressure is increased beyond 43 GPa [38]. Similarly, the previous work on BF-0.35PT also did not observe the superlattice peaks characteristic of the presumed AFD transition [52]. The observation of an AFD transition at a pressure $p_c$ ~2.15 GPa, which is lower than that predicted by first principles calculations for $PbTiO_3$ [33], confirms that BF substitution enhances the AFD instability of $PbTiO_3$ at a lower pressure. This is not surprising as BF has a very strong AFD instability [41-42].

We now proceed to determine the structure of the high pressure phase resulting from the tetragonal phase at $p_c$ ~2.15 GPa. First principles calculations by Frantti et al. [35] on $PbTiO_3$ predict a tetragonal P4mm to R3c transition in $PbTiO_3$. In case of PTBF50, the doublet nature of $(220)_{pc}$ and $(222)_{pc}$ and singlet character of $(200)_{pc}$ may appear to suggest rhombohedral structure in the R3c space group at $p_c$ ~2.15 GPa . However, careful profile analyses using peak deconvolution, Le Bail refinement and Rietveld refinement prove it otherwise. To demonstrate this, we first present the results of peak deconvolution of the $(220)_{pc}$ and $(222)_{pc}$ profiles using two pseudo-Voigt functions of equal widths, as expected from Caglioti relationship for peak width (FWHM = u*tan$^2\theta$ + v*tan$\theta$ + w), for neighbouring reflections [53]. The results of peak deconvolution are shown in Figs. 3 and 4, which reveal huge mismatch between the observed and fitted profiles for the rhombohedral structure. This gave us the first indication that the high pressure phase appearing at $p_c$ ~2.15 GPa may not be rhombohedral. Cohen and his co-workers have predicted theoretically and verified experimentally the occurrence of tetragonal P4mm to monoclinic Pm to another monoclinic Cm phase transitions in PbTiO$_3$ involving rotation of the ferroelectric polarization vector [26, 36-37]. In our case, because of the observation of the superlattice peaks, the space group of the monoclinic phase would change as reported in the context of the low temperature phase of PZT ceramics, where the space group changes from Cm to Cc [31]. The composition versus temperature phase diagram of PT-xBF system at ambient pressure shows phase transition from tetragonal P4mm to monoclinic Cc phase with increasing BF-content passing via a morphotropic phase boundary (MPB) region supporting polarization rotation [47]. For the monoclinic Cc phase, the $(220)_{pc}$ and $(222)_{pc}$ are expected to be quadruplet and triplet, respectively. In view of this, we decided to use more than two peaks of equal widths for fitting the $(220)_{pc}$ and $(222)_{pc}$ profiles. We find that the $(222)_{pc}$ profile can be fitted satisfactorily using three peaks of equal width as shown in Fig. 3. For the $(220)_{pc}$ profile, we attempted both 3 and 4 peaks of equal widths and the results are shown in Fig. 4. It is evident from these figures that the fit between the observed and fitted profiles for $(220)_{pc}$ is unsatisfactory even after using three peaks of equal width. However, we obtained an excellent fit on considering four peaks of equal width. Thus, the peak deconvolution analysis of the $(222)_{pc}$ and $(220)_{pc}$ profiles in terms of three and four peaks, respectively, suggests that the structure of the high pressure phase at ~2.15 GPa may be monoclinic.

To determine the space group of the high pressure phase, we first carried out profile matching analysis using LeBail method for the data collected at p ~2.15 GPa. For LeBail refinement, we considered the plausible isotropy subgroups of the tetragonal P4mm space group resulting from freezing of the optical zone centre ($\Gamma_4^-$ mode, q = 0,0,0) and zone boundary ($R_4^+$ mode, q = ½,½,½) phonon modes. The ISOTROPY software suite [54] predicts R3c, Fmm2, Ima2, Imm2, I4cm, Cm, C2, Cc and P1 space groups as the plausible space groups [47]. All these possible space groups were considered for the profile matching analysis using LeBail technique, except for the triclinic P1 space group which is the least symmetric. The results of the LeBail refinement using the SXRD data for all these space groups are shown in Fig. 5 and Fig. S2 of the supplementary information [55]. The refinement for the monoclinic Cc space group was carried out in the Ic setting, which is crystallographically equivalent to the conventional Cc setting. The Ic setting allows us to visualise the relationship between the monoclinic and the elementary perovskite unit cell parameters conveniently as was done in the context of PZT also [27]. It is evident from the difference profile shown in Fig. S2 [55] that the tetragonal (I4cm), orthorhombic (Fmm2, Ima2 and Imm2) and rhombohedral R3c space groups do not give satisfactory fits, particularly for $(hhh)_{pc}$ type reflections. Among the monoclinic space groups Cm, C2 and Ic(≡Cc), the Ic(≡Cc) space group gave the best fit for the main perovskite as well as the superlattice reflections as can be seen from Fig. 5 and S2 of the supplementary information [55]. The value of $R_{wp}$ is lower for the monoclinic Ic space group ($R_{wp}$ = 3.23%) than that for the rhombohedral R3c space group ($R_{wp}$ = 5.20%) and other space groups. The LeBail refinement thus suggests that the space group of the high pressure phase of PTBF50 at p ~2.15 GPa is monoclinic Ic(≡Cc). In view of the very good quality of the data, we also performed Rietveld refinements using our high pressure SXRD data considering both the rhombohedral R3c and monoclinic Ic(≡Cc) space groups. The fit between the observed and calculated profiles for both the space groups are shown in Fig. 6. It is evident from these figures that Ic(≡Cc) space group gives far better fit with lower $R_{wp}$ and $\chi^2$ ($R_{wp}$ = 10.9%, $\chi^2$= 1.40) as compared to that for the R3c space group ($R_{wp}$ = 24.5%, $\chi^2$= 7.56). Thus both, the Le Bail and Rietveld refinements, show that the high pressure phase of PTBF50 for p ~2.15 GPa is monoclinic having Ic(≡Cc) space group. This space group corresponds to the $a^-a^-c^-$ tilt system in the Glazer's notation for perovskites with tilted oxygen octahedra [50-51] with superimposed ferroelectric distortion, as reported in the case of PZT [27,56]. This result on the pressure induced P4mm to Ic(≡Cc) transition, on the one hand, supports the theoretical predictions about the existence of AFD transition by Kornev and his co-workers [33-34] at moderate pressures, it also supports the polarization rotation model, on the other, as the ferroelectric polarization lies on a symmetry plane, predicted by Cohen and his co-workers [26,36]. However, Kornev and his co-workers predict I4cm space group, while Cohen and his co-workers predict Cm space group without AFD transition, none of which gives the best fit, as can be seen from Fig. S2 of the supplementary information [55].

Having confirmed the Ic(≡Cc) space group for the high pressure phase of PTBF50 at $p_c$ ~2.15 GPa, LeBail refinement was carried out for various high pressure diffraction patterns up to ~12.61 GPa using Ic setting of the Cc space group. Fig. 7 depicts the variation of pseudo-cubic lattice parameters and the monoclinic distortion angle β, so obtained from the LeBail refinements, as a function of pressure. The pseudo-cubic lattice parameters were obtained from the Ic unit cell parameters using the relationships $a_{Ic} \simeq a_{pc}\sqrt{2}$, $b_{Ic} \simeq b_{pc}\sqrt{2}$, and $c_{Ic} \simeq 2c_{pc}$ [27]. The monoclinic angle β of the Ic unit cell is same as that of the pseudo-cubic unit cell. It can be seen from Fig. 7(a) that pressure reduces the '$c_{pc}$' unit cell parameter quite drastically, whereas the increase in the '$a_{pc}$' parameter is rather modest in the tetragonal phase field. The reduction in the '$c_{pc}$' parameter shows that the system is trying to accommodate pressure by reducing the ferroelectric distortion in the c-direction as expected on the basis of Samara's criterion [24]. After the initial rising trend of the '$a_{pc}$' parameter with increasing pressure in the tetragonal phase field, '$a_{pc}$' along with the '$b_{pc}$' and '$c_{pc}$' lattice parameters show a monotonic decrease in the monoclinic phase field for p ≥ 2.15 GPa. This trend of first increase of the '$a_{pc}$' at lower pressures followed by its decreasing behaviour along with continuously decreasing behaviour of the '$c_{pc}$' lattice parameter is similar to that reported by Nelmes et al. for pure PT, but they could not identify the monoclinic distortion [49]. Interestingly, the inequality relationship in the $a_{pc}$, $b_{pc}$ and $c_{pc}$ parameters of the monoclinic phase changes from $a_{pc} > b_{pc} \sim c_{pc}$ for ~2.15 ≤ p ≤ 7 GPa to $a_{pc} < b_{pc} \sim c_{pc}$ for p >7 GPa. It was verified by LeBail and Rietveld refinements that the structure remains monoclinic in the Ic(≡Cc) space group even after 7 GPa. The signature of this reversal in the relative values of the cell parameters is clearly seen in the $(222)_{pc}$ profile also (see Fig. 1), where the weaker intensity peak on the lower 2θ side for p <7 GPa occurs on the higher 2θ side for p >7 GPa. This reversal in the relative values of the pseudo-cubic lattice parameters in the monoclinic region is similar to the $M_A$ to $M_B$ type isostructural phase transition, where both $M_A$ and $M_B$ belong to the same space group, theoretically predicted by Cohen & Vanderbilt [43] and verified in PMN-xPT ceramics as a function of composition [57] with one important difference. In the PMN-xPT, there is freezing of the zone centre ferroelectric mode only, whereas in PTBF50 system, both ferroelectric and AFD modes are frozen. As a result, the Cm space group of PMN-xPT [57] changes to Ic(≡Cc) discovered first in PZT at low temperatures in the MPB region [31]. The isostructural phase transition at $p_{c2} \sim$ 7 GPa is most clearly revealed in the variation of the monoclinic angle β with pressure, shown in the inset (i) of Fig. 7(a) which shows a minimum at $p_{c2}$ ~7 GPa. The monoclinic angle β initially decreases with increasing pressure up to $p_{c2}$ ~7 GPa and approaches a value close to 90 degree, after which it starts increasing again for p >7 GPa. This clearly suggests a phase transition around $p_{c2} \simeq 7$ GPa.

The signature of the P4mm to Ic(≡Cc) of $M_A$ type followed by Ic(≡Cc) of $M_A$ type to Ic(≡Cc) of $M_B$ type phase transitions at $p_{c1}$ ~2.15 GPa and $p_{c2}$ ~7 GPa, respectively are also seen in the equation of state (p-V) shown in Fig. 7(b), where V is the unit cell volume per perovskite formula unit for all the structural phases. The sharp discontinuity in V at $p_{c1}$ ~2.15 GPa reveals first order character of the tetragonal P4mm to monoclinic Ic(≡Cc) phase transition. For the second transition, there is a very weak signature in the unit cell volume vs. linear pressure plot suggesting it to be weakly first order or nearly second order type. When we plot the unit cell volume against log of pressure (log p) as shown in the inset of Fig. 7(b), a deviation from the linearity can be seen on the higher pressure side at $p_{c2}$ ~7 GPa, which is the pressure at which the reversal in the inequality relationship of the lattice parameters is also observed and the monoclinic distortion angle β shows a sharp dip (see Fig. 7(a) and the inset (i) of Fig. 7(a)). All these observations clearly suggest the presence of another phase transition, albeit isostructural, at $p_{c2} \simeq 7$ GPa. We designate the monoclinic phase for ~2.15≤ p<7 GPa and p ≥7 GPa pressure regions as Ic-I(≡Cc-I) and Ic-II(≡Cc-II), respectively. This type of pressure induced isostructural phase transition involving change in the inequality relationship of lattice parameters has been reported in case of BiOCl also [58]. The equation of state in the stability fields of the two monoclinic phases and for the tetragonal phase below p < $p_c$ ~2.15 GPa, was fitted to third order Birch-Murnaghan model [59] using EOSFIT software [60]. The bulk modulus 91.1 ± 7.3 GPa of the Ic-I(≡Cc-I) phase so obtained is nearly equal to the average of reported bulk moduli of $BiFeO_3$ (75.5 ± 15.5 GPa) [61] and $PbTiO_3$ (100 ± 7 GPa) [62] mixed in 1:1 ratio. The bulk modulus for the Ic-II(≡Cc-II) phase was obtained to be 105.1 ± 3.9 GPa, which is higher than the previous phase indicating lower compressibility of the system at higher pressures, as anticipated due to the assertion of short range repulsive interactions.

We now address the issue of the so-called cubic like phase observed at the intermediate pressures in PbTiO$_3$ by various workers [33,38]. Similar to PbTiO$_3$, the XRD profiles of PTBF50 also do not exhibit any splitting for the pressure range ~5 ≤ p ≤ 7 GPa, giving the impression as if the structure has become cubic. The cubic like intermediate phase reported by Janolin et al. [38] in pure PT was assumed to be either an average cubic phase in the Pm$\bar{3}$m space group with local distortions or tetragonal I4cm phase without any LeBail and/or Rietveld refinements. In case of BF-0.35PT, it has been assigned a non-ferroelectric cubic Pm$\bar{3}$m space group by Basu et al. [52]. However, the presence of superlattice reflections (see Fig. 2) in this region reveals that the structure is still antiferrodistorted and therefore the possibility of cubic Pm$\bar{3}$m phase corresponding to the elementary perovskite cell can be ruled out straight away. The only non-ferroelectric cubic space group, as per Glazer's classification of the tilted octahedral structures, is Im3 with doubled pseudo-cubic unit cell parameters [50-51]. But it can also be rejected as it corresponds to all in-phase tilts with $a^+a^+a^+$ tilt system in Glazer's notation which should have resulted in superlattice reflections with odd-odd-even type of indices [50-51] which is clearly not the case here. The superlattice reflections present in the 'cubic' like phase of PTBF50 have (311)$_{pc}$ and (332)$_{pc}$ type indices, with respect to the doubled perovskite unit cell, which are odd-odd-odd type and not odd-odd-even type [50-51]. In this cubic-like region, both Ic(≡Cc) and R3c space groups gave nearly comparable profile fits. However, there is no signature of any phase transition in the pressure region ~5 ≤ p < 7 GPa in the equation of state. Therefore we can safely conclude that the so-called cubic like phase of PTBF50 is the antiferrodistorted monoclinic phase Ic-I(≡Cc-I) itself with very small monoclinic distortion but with significant tilting of oxygen octahedra. The variation of the monoclinic angle β also does not give any signature of another phase transition in this intermediate pressure region as it decreases monotonically with $\beta \rightarrow 90^o$ at $p_{c2}$ ~7 GPa. A pressure induced rhombohedral phase between two monoclinic regions of lower symmetry is quite unlikely based on symmetry considerations also. The R3c space group is not a subgroup of Ic($\equiv$Cc) and, as such, Ic($\equiv$Cc) to R3c transition has to be a first order phase transition [63], for which a discontinuous change of unit cell parameters is expected. Since no such discontinuity is observed in this intermediate pressure region in the equation of state shown in Fig. 7(b), we are led to propose that the cubic-like feature in the diffraction patterns correspond to the crossover regime of isostructural phase transition from the Ic-I($\equiv$Cc-I) of $M_A$ type to Ic-II($\equiv$Cc-II) phase of $M_B$ type.

We finally address the most contentious issue of the presence of the reentrant ferroelectric phase at high pressures, predicted by Kornev and his co-workers [33-34]. Since the data quality is very good, as noted by other workers also for the pressure dependent SXRD data collected on the same beamline under similar experimental conditions [64-65], Rietveld refinement was performed in order to get an idea about the variation of the ferroelectric polarization and the oxygen octahedral tilting as a function of pressure. The ferroelectric polarization was calculated using the refined positional coordinates and the Born effective charges for BF [42] and PT [40] mixed in 1:1 ratio in the manner discussed in Ref. [66]. Since the structure of PTBF50 in the Ic($\equiv$Cc) space group possesses ferroelectric distortion as well as tilting of oxygen octahedra, it is not possible to calculate the exact tilt angles from the refined coordinates. Octahedral tilting bends the Fe/Ti-O-Fe/Ti bonds due to the change of oxygen position in a direction nearly perpendicular to the <100>$_{pc}$ direction. We have used one of the Fe/Ti-O-Fe/Ti bond angles in the equivalent pseudo-cubic unit cell, which is the least affected by the ferroelectric distortion, to get the trend of the change in the tilt angle as a function of pressure. The variation of the ferroelectric polarization and the oxygen octahedral tilt angle so obtained are shown in Figs. 8(a) and (b). In the tetragonal P4mm region, external hydrostatic pressure drastically decreases the ferroelectric polarization as anticipated from Samara's criterion also [24]. At $p_{c1}$ ~2.15 GPa, the ferroelectric polarization was found to exhibit a drastic increase corresponding to the phase transition to the monoclinic Ic($\equiv$Cc) phase. This is akin to the sharp increase in ferroelectric polarization in well-known piezoelectric ceramics like PZT at the MPB on varying the composition which changes the structure from tetragonal to monoclinic [27]. In our case, the MPB-like effect is induced not by chemical pressure (like bigger $Zr^{4+}$ ion in place of smaller $Ti^{4+}$ ion in $PbTiO_3$), but by external pressure. This observation further supports the theoretical predictions about the pressure induced rotation of the polarization vector and MPB effect in pure $PbTiO_3$ by Cohen and his co-workers [26,36]. After the initial jump in the ferroelectric polarization at the critical pressure $p_{c1}$ ~2.15 GPa, it starts decreasing with further increase of pressure, whereas the tilt angle does not exhibit any remarkable change with increasing pressure. This indicates that pressure is getting accommodated mainly by decrease in the polarization in this regime as per Samara's criterion [24]. However, the ferroelectric polarization starts increasing again after the Ic-I($\equiv$Cc-I) to Ic-II($\equiv$Cc-II) isostructural phase transition at $p_c \sim$ 7 GPa, revealing the presence of the reentrant ferroelectric phase. Kornev et al. [54,55] have predicted theoretically that the increase in the ferroelectricity of pure $PbTiO_3$ and other similar compounds in the reentrant ferroelectric phase occurs to reduce the overlap of Ti 3d and O 2s orbitals coming closer to each other with increasing pressure. We find that the high pressure is getting accommodated efficiently in this region through the steep increase in the oxygen octahedral tilt angle shown in Fig. 8(b). Thus, pressure induced volume reduction is achieved primarily through decrease in the ferroelectric polarization and increase in the oxygen octahedral tilt angle in the Ic-I(Cc-I)and reentrant Ic-II(Cc-II) phase fields, respectively. We believe that the complementary nature of the pressure dependence of ferroelectric polarization and oxygen octahedral tilting within the Ic-I(Cc-I) and Ic-II(Cc-II) phase fields, observed in our study, provides an efficient mechanism for accommodating high pressure induced volume reduction.

## IV. CONCLUDING REMARKS

In the present work, we have discovered new high pressure phase transitions in a tetragonal composition of BiFeO$_3$ (BF) substituted PbTiO$_3$ (PT) in the solid solution system (Pb$_{(1-x)}$Bi$_x$)(Ti$_{(1-x)}$Fe$_x$)O$_3$ with x = 0.50 (PTBF50) using high resolution synchrotron x-ray diffraction (SXRD) studies. BF substitution was selected to enhance the antiferrodistortive (AFD) instability of PbTiO$_3$ so that the weak superlattice reflections arising from the theoretically predicted AFD transition in ferroelectric tetragonal PbTiO$_3$ become discernible in the SXRD patterns. The results presented in this paper amply justify this strategy as the AFD superlattice reflections have been observed by us in our SXRD diffraction patterns at moderate pressures $p_c$ ~2.15 GPa for the first time. The direct observation of these superlattice peaks has enabled us to get a full picture of the high pressure phase transition behaviour of tetragonal PTBF50 with the help of which we have successfully addressed all the existing controversies about the high pressure behaviour of the tetragonal phase of pure PbTiO$_3$. Our main findings can be summarised as follows: (1) We have shown that the tetragonal phase of PTBF50 transforms to a monoclinic phase at a critical pressure of $p_{c1}$ ~2.15 GPa. This transition involves not only the rotation of the ferroelectric polarization vector as predicted theoretically by Cohen and his co-workers [26,36] for PT, but is also accompanied with a concomitant AFD transition predicted by Kornev and his co-workers [33-34]. Analysis of the SXRD data, including the weak superlattice reflections which are characteristic of the AFD transition, using careful peak deconvolution, LeBail refinement and Rietveld refinement shows that the structure of the higher pressure phase of PTBF50 corresponds to the monoclinic Cc (or equivalently Ic in the non-standard setting) and not the R3c space group proposed in an earlier work [52]. (2) The ferroelectric polarization, obtained from the structural data and Born effective charges, is found to take a sharp jump at this tetragonal (P4mm) to monoclinic (Ic) phase transition pressure $p_{c1}$ ~2.15 GPa, reminiscent of a similar peak in the technologically important piezoelectric ceramics at the MPB composition [21]. (3) With further increase in pressure, the ferroelectric polarization of the monoclinic phase is found to decrease, as expected from Samara's criterion also [24], while the octahedral tilt angle remains nearly unaffected. (4) At around $p_{c2}$ ~7 GPa, the monoclinic phase Ic-I(≡Cc-I) undergoes an isostructural phase transition to another monoclinic Ic-II(≡Cc-II) phase such that the $a_{pc}>b_{pc}\simeq c_{pc}$ with decreasing trend of monoclinic angle β of the elementary perovskite unit cell of the lower pressure Ic-I(≡Cc-I) phase changes to $a_{pc}<b_{pc}\simeq c_{pc}$ with β increasing in the higher pressure phase Ic-II(≡Cc-II). These two orientations of the monoclinic phase are reminiscent of the $M_A$ and $M_B$ phases in the theoretically predicted phase diagram of Vanderbilt and Cohen for PZT and related systems [43]. (5) We have also shown that the Ic-II(≡Cc-II) phase is the reentrant ferroelectric phase, predicted theoretically by Kornev and Bellaiche [33-34] in the context of $PbTiO_3$, but never observed experimentally, in which the ferroelectric polarization shows an anomalous increase with pressure, contrary to Samara's criterion [24]. Furthermore, this increase in polarization with pressure is facilitated by a concomitant increase in the oxygen octahedral tilt angle which ensures the volume compression via Ti/Fe-O-Ti/Fe bond bending and not the Ti/Fe-O bond compression considered in Samara's criterion [24]. (6) The so-called "cubic"-like phase, experimentally reported in $PbTiO_3$ [38] and BF-0.35PT [52], is shown to have the same AFD structure, with characteristic superlattice peaks, as the two isostructural monoclinic phases. The "cubic-like" feature of the main perovskite reflections is shown to be linked with the crossover from the ferroelectric Ic-I(≡Cc-I) to Ic-II(≡Cc-II) isostructural phase transition.

The present work on the pressure induced phase transitions in tetragonal PTBF50 not only addresses the various controversies related to the pressure induced phase transitions of pure

PbTiO$_3$, it also provides us a clue to understand the origin of the morphotropic phase boundary at ambient pressures in the perovskite systems. In the case of PT-xBF solid solution system, BF has a smaller unit cell volume and its substitution in PbTiO$_3$ is expected to generate chemical pressure. We believe that it is this chemical pressure that drives the tetragonal P4mm phase of PbTiO$_3$ to transform to the monoclinic Cc phase at ambient pressures in the PT-xBF solid solution system [47] much in the same way as external critical pressure p$_c$ ~2.15 GPa transforms the tetragonal PTBF50 to a similar monoclinic structure. A crude estimate, based on the linear interpolation of the unit cell volumes and bulk moduli of BF and PT, shows that the chemical pressure generated by BF in the PT-xBF system is of the order of 0.5 GPa for the MPB composition which is quite close to the experimental critical pressure p$_c$ ~2.15 GPa at which PTBF50 transforms to monoclinic Cc phase under external pressure at ambient temperatures. The understanding of the origin of the MPB in the PT-xBF system and the existence of the reentrant ferroelectric phase at high pressures provides an insight into designing new eco-friendly lead-free piezoelectric systems having MPB-like characteristics with enhanced electromechanical coupling via introduction of chemical pressure. The required ingredients for designing such a system are - suitable base ferroelectric perovskite system and appropriate dopant(s) that can not only generate sufficient chemical pressure but also favour tilting of the octahedral required for bending of the chemical bonds to accommodate the pressure for the reentrant ferroelectric phase to be stabilised. We believe that by properly playing with the zone centre and zone boundary optical phonon instabilities, new environmentally friendly non-toxic lead free ferroelectric systems can be designed with piezoelectric properties better or at least comparable to that of present day lead-containing MPB-based solid solution systems.

## ACKNOWLEDGEMENTS

P.S. acknowledges support from the Department of Science and Technology (DST), Government of India, through the award of DST INSIRE Fellowship. Portions of this research were carried out at the light source PETRA III of DESY, a member of the Helmholtz Association (HGF). Financial support by the Department of Science & Technology (Government of India) provided within the framework of India@DESY collaboration, operated through Saha Institute of Nuclear Physics (Kolkata)/Jawaharlal Nehru Centre for Advanced Scientific Research (Jakkur), India, is gratefully acknowledged.

**Figure captions:**

FIG. 1. Evolution of $(200)_{pc}$, $(220)_{pc}$ and $(222)_{pc}$ perovskite profiles of PTBF50 with increasing pressure.

FIG. 2. Evolution of $(3/2\ 1/2\ 1/2)_{pc}$ {denoted by *} and $(3/2\ 3/2\ 1/2)_{pc}$ {denoted by #} superlattice peaks due to AFD transition in PTBF50 with increasing pressure.

FIG. 3. Peak deconvolution of $(222)_{pc}$ profile at p ~2.15 GPa using two and three peaks of equal width. The clear mismatch between the observed (open circles) and fitted profile (solid line) near the tail region of the profile shows that it cannot be fitted accurately using only two peaks.

FIG. 4. Peak deconvolution of $(220)_{pc}$ profile at p ~2.15 GPa using two, three and four peaks of equal width. The clear mismatch between the observed (open circles) and fitted profile (solid line) near the tail region of the profile shows that it cannot be fitted accurately using only two or three peaks.

FIG. 5. Observed (red open circles), calculated (black continuous line) and difference (bottom blue line) profiles obtained from LeBail refinement at p = 2.15 GPa using (a) R3c and (b) Ic(Cc) space groups at ambient temperature. The vertical lines (pink) indicate the Bragg peak positions.

FIG. 6. Observed (red open circles), calculated (black continuous line) and difference (bottom blue line) profiles obtained from Rietveld refinement at p = 2.15 GPa using (a) R3c and (b) Ic(Cc) space groups at ambient temperature. The vertical lines (pink) indicate the Bragg peak positions.

FIG. 7. Variation of (a) pseudocubic (elementary perovskite) unit cell parameters $a_{pc}$ (red filled circle), $b_{pc}$ (blue '+' symbol) and $c_{pc}$ (open black circle), and (b) unit cell volume $V_{pc}$ as a function of pressure. Insets (i) and (ii) of (a) show variation of monoclinic angle β and magnified view of $a_{pc}$, $b_{pc}$ and $c_{pc}$ with pressure. Inset of (b) shows variation of $V_{pc}$ with log(pressure).

FIG. 8. Variation of (a) ferroelectric polarization and (b) oxygen octahedral tilt angle (tilting of one of the B-O-B bond angles) as a function of pressure.

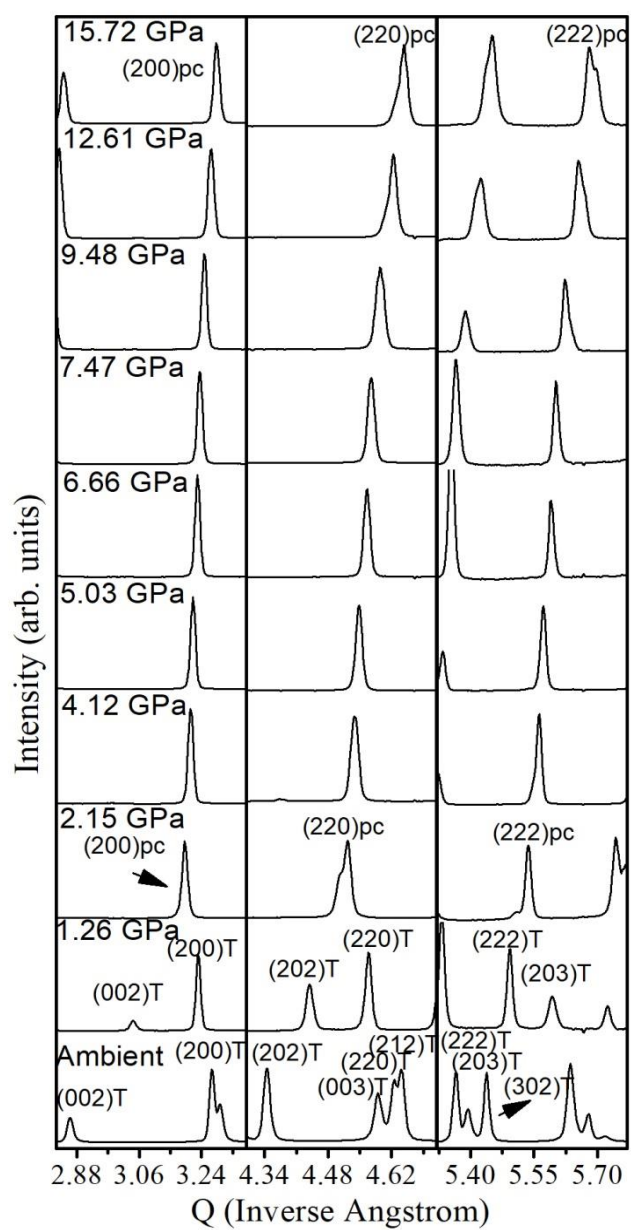

Figure 1.

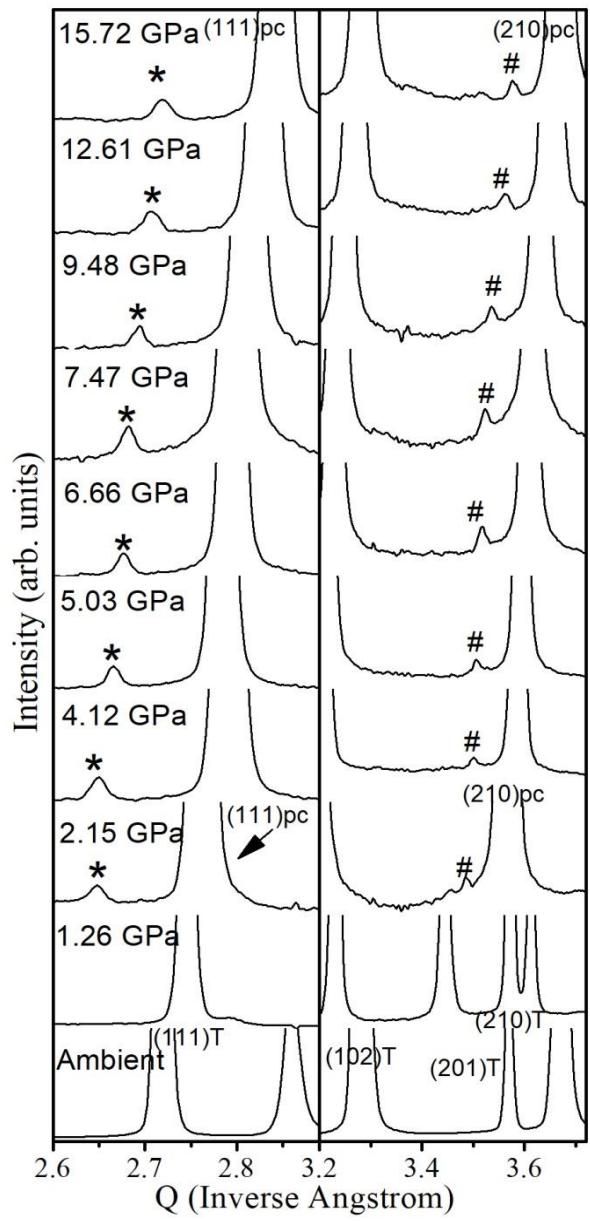

Figure 2.

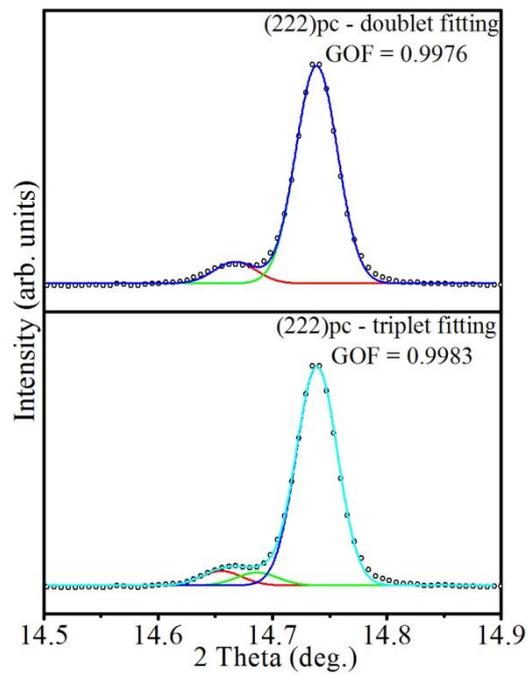

Figure 3.

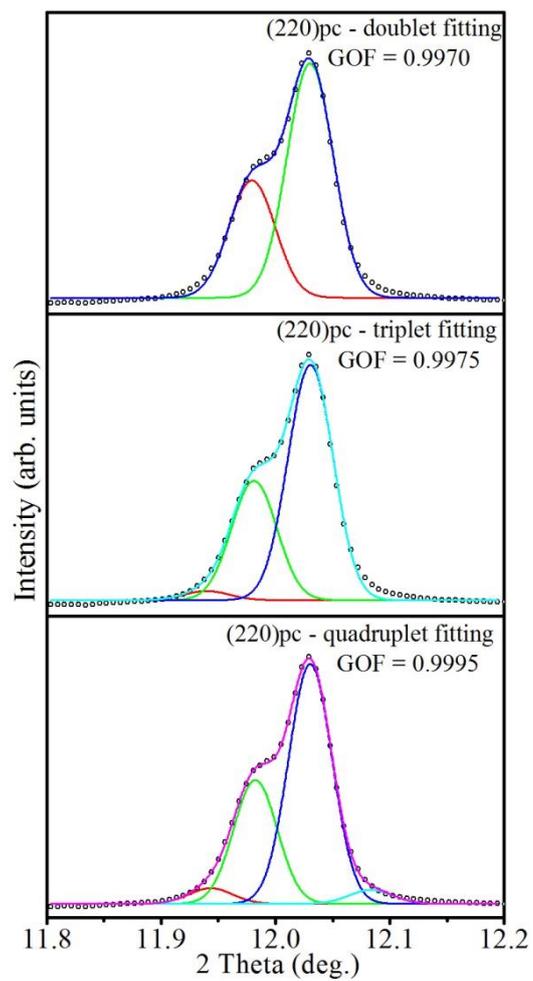

Figure 4.

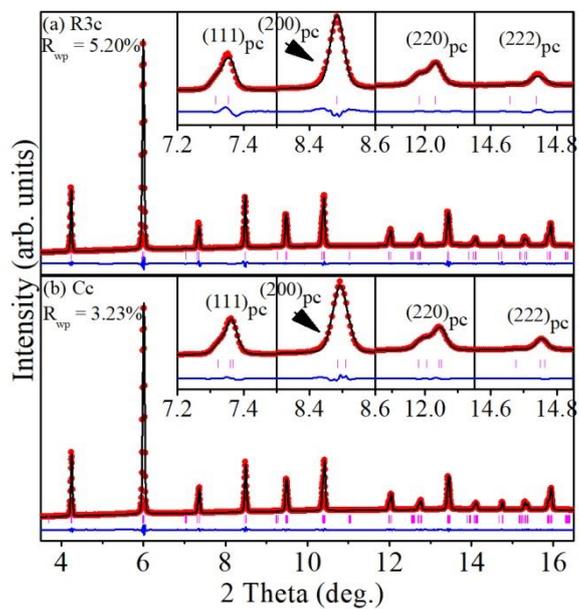

Figure 5.

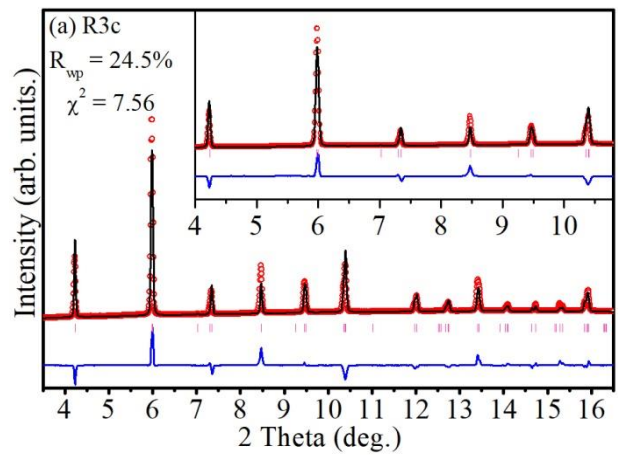

Figure 6(a).

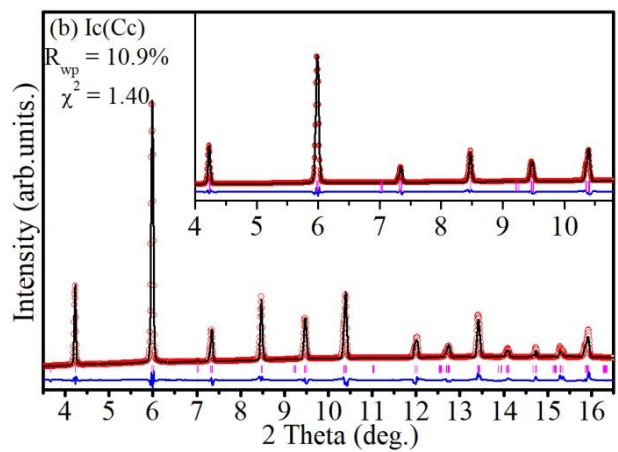

Figure 6(b).

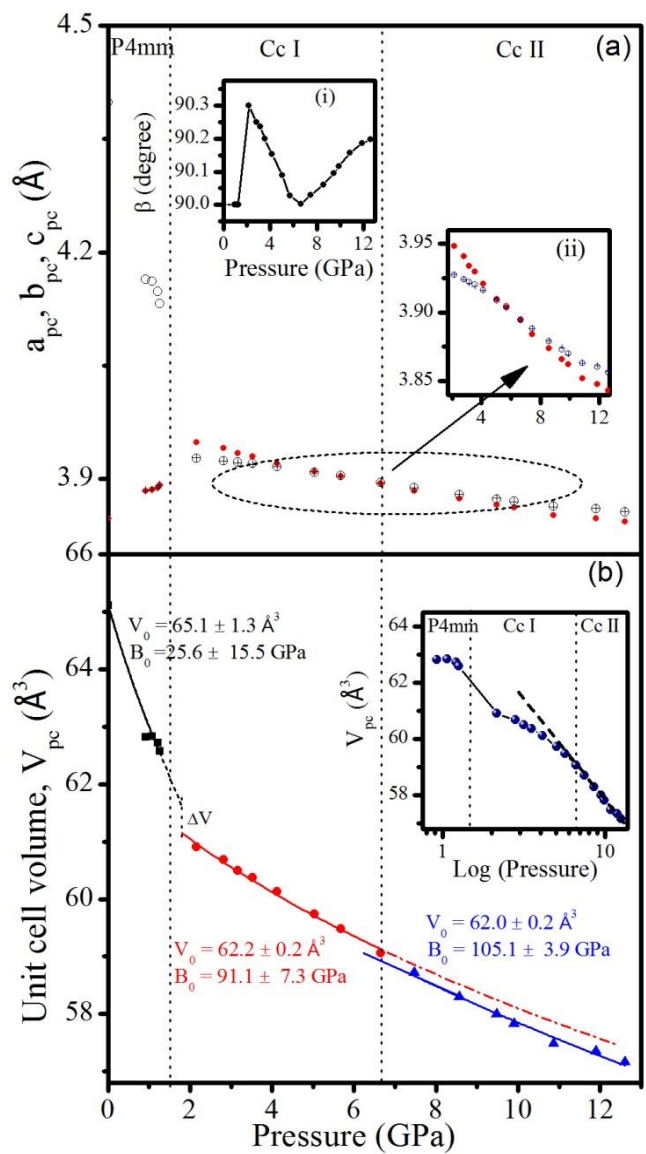

Figure 7.

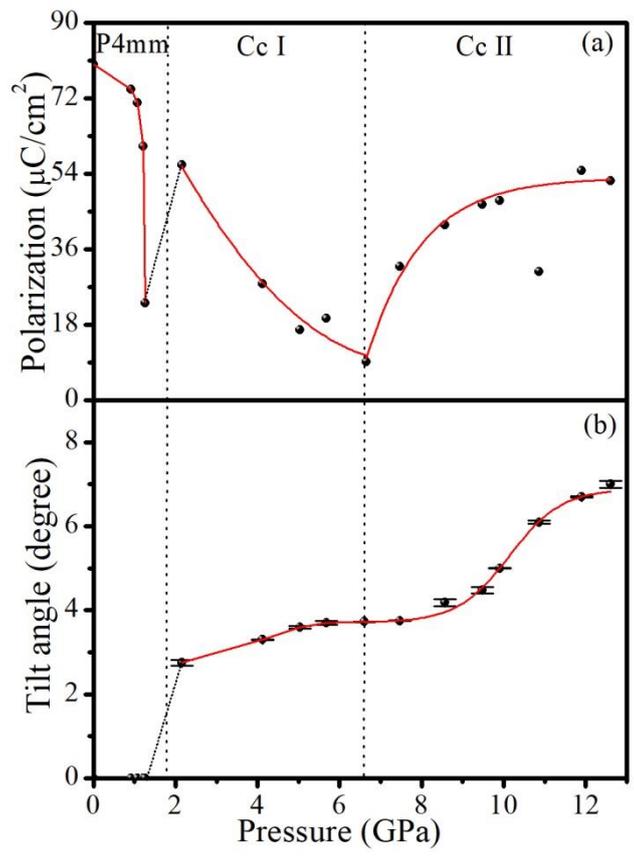

Figure 8.